\documentclass[aps,twocolumn,superscriptaddress,showpacs]{revtex4}
\usepackage{epsfig}
\usepackage{graphicx}
\begin{document}
\newcommand{\tdna}{t^{\dagger}_{n\alpha}}
\newcommand{\tna}{t_{n\alpha}}
\newcommand{\tdnoa}{t^{\dagger}_{n+1,\alpha}}
\newcommand{\tnoa}{t_{n+1,\alpha}}
\newcommand{\tdng}{t^{\dagger}_{n\gamma}}
\newcommand{\tng}{t_{n\gamma}}
\newcommand{\tdnod}{t^{\dagger}_{n+1,\delta}}
\newcommand{\tnod}{t_{n+1,\delta}}
\newcommand{\tdag}{t^{\dagger}}
\newcommand{\al}{\alpha}
\newcommand{\be}{\beta}
\newcommand{\ca}{\gamma}
\newcommand{\de}{\delta}
\newcommand{\taud}{\tau^{\dagger}}
\draft
\title{\bf A modified triplet-wave expansion method applied to the
alternating Heisenberg chain}
\author{ A. Collins, C. J. Hamer and Zheng Weihong}
\affiliation{School of Physics, The University of New South Wales,
  Sydney, NSW 2052, Australia}
\date{\today}
\begin{abstract}
An alternative triplet-wave expansion formalism for dimerized spin systems
 is presented, a modification of the `bond
operator' formalism of Sachdev and Bhatt. Projection operators are used to confine
 the system to the physical
subspace, rather than constraint equations. The method is
illustrated for the case of the alternating Heisenberg chain, and
comparisons are made with the results of dimer series expansions
and exact diagonalization. Some discussion is included of the
phenomenon of ``quasiparticle breakdown", as it applies to the
two-triplon bound states in this model.
\end{abstract}
\pacs{PACS Indices: 05.30.-d,75.10.-b,75.10.Jm,,75.30.Kz \\
\\  \\
(Submitted to  Phys. Rev. B) }
\maketitle
\newpage

%\narrowtext
\section{INTRODUCTION}
\label{sec1}

There has been much interest recently in the phenomenon of {\it dimerization}
 in $S=1/2$ Heisenberg antiferromagnets, where pairs of neighbouring
spins couple to form $S=0$ singlet dimers. The dimerization may arise due to
inhomogeneous bond interactions, as in the alternating Heisenberg chain (AHC)
 model, or the Shastry-Sutherland model in two dimensions \cite{shastry1981}.
 Alternatively, it
may emerge spontaneously, as the result of frustration
\cite{lhuillier2001}: this seems to occur in the J$_1$-J$_2$ square lattice model at
 intermediate coupling values, for instance,
although there is disagreement as to whether
the pattern of dimerization is ordered (`valence bond solid')
 \cite{read1991,kotov1999} or disordered (`valence bond liquid' or
`resonating valence bond') \cite{anderson1987,capriotti2003}.

To understand the properties of dimerized phases, it is useful to construct an
 appropriate lattice formalism describing the dimers and their spin-triplet excitations.
 The
physics of the system can then be connected with the properties of the elementary
triplet excitations; and one can also use the formalism to construct a continuum
`effective Lagrangian' field theory for the system at hand. Such a formalism was the
`bond-operator' representation constructed by Sachdev and Bhatt
\cite{sachdev1990} (see also Chubukov \cite{chubukov1989}) some years ago, which is
analogous to the spin-wave representation traditionally used to describe the magnetically
ordered phases of these systems \cite{mattis1981}.

Sachdev and Bhatt \cite{sachdev1990} considered two spins ${\bf S_1}$ and ${\bf S_2}$ at
 either end of a single bond on the lattice, forming a dimer. They introduced a singlet
 and
three triplet boson creation operators to form the corresponding states from the vacuum:

\begin{eqnarray}
|s> & = & s^{\dagger} |0> = \frac{1}{\sqrt{2}}(|\uparrow \downarrow> - | \downarrow
 \uparrow>) \nonumber \\
|t_x> & = & t^{\dagger}_x |0> = -\frac{1}{\sqrt{2}}(|\uparrow \uparrow> - | \downarrow
 \downarrow>) \nonumber \\
|t_y> & = & t_y^{\dagger} |0> = \frac{i}{\sqrt{2}}(|\uparrow \uparrow> + | \downarrow
 \downarrow>) \nonumber \\
|t_z> & = & t_z^{\dagger} |0> = \frac{1}{\sqrt{2}}(|\uparrow \downarrow> + | \downarrow
 \uparrow>)
\label{eq1}
\end{eqnarray}

Then the spin operators can be represented

\begin{eqnarray}
S_{1\alpha} & = & \frac{1}{2}[s^{\dagger}t_{\alpha} + t_{\alpha}^{\dagger}s
-i\epsilon_{\alpha \beta \gamma}t^{\dagger}_{\beta}t_{\gamma}]
\nonumber \\
S_{2\alpha} & = & \frac{1}{2}[-s^{\dagger}t_{\alpha} - t_{\alpha}^{\dagger}s
-i\epsilon_{\alpha \beta \gamma}t^{\dagger}_{\beta}t_{\gamma}]
\label{eq2}
\end{eqnarray}
(where $\alpha,\beta,\gamma$ take values $x,y$ or $z$), with the constraint that physical
 states must satisfy

\begin{equation}
s^{\dagger}s + t^{\dagger}_{\alpha}t_{\alpha} = 1.
\label{eq3}
\end{equation}
They applied this formalism to develop a mean field theory of the frustrated
square-lattice antiferromagnet.

The problem with this approach is that the constraint (\ref{eq3}) is awkward to implement
 analytically. Kotov {\it et al.} \cite{kotov1998} have applied an alternative
``Brueckner approach", in which the singlet operator is discarded, leaving only the constraint that two
 triplet excitations are not allowed on the same site (bond). This
is implemented by an infinite on-site repulsion term between triplets, which is applied using an analytic
 Brueckner approach, valid when the density of triplets is
small. The approach has been applied to the two-layer Heisenberg model \cite{kotov1998,shevchenko1999},
the quantum
 spin-ladder \cite{sushkov1998,kotov1999a}, and the dimerized Heisenberg chain with frustration
\cite{shevchenko1999a}, and some useful physical insights have been obtained. In particular, the
 occurrence of two-particle bound states formed from the elementary triplet excitations
seems to be generic in these models. Nevertheless, the Brueckner implementation of the on-site repulsion
 term is also somewhat awkward to apply, and difficult to carry
through in higher orders.

Here we present an alternative approach in which the triplet exclusion constraint is implemented
 automatically by means of projection operators. We also use a
``modified" formalism, analogous to modified spin-wave theory \cite{takahashi1987,gochev1994}, in which
 the two-body terms in the Hamiltonian are diagonalized through to the highest order
calculated. The absence of any constraint makes the formalism easier and more transparent to apply. The
 only drawback is the appearance of extra many-body interaction
terms in the Hamiltonian, so that carrying the calculation to high orders would require the aid of a
 computer.

To illustrate the formalism, we apply it to the case of the alternating Heisenberg chain (AHC). This model has itself attracted much attention recently, as new materials
such as Cu(NO$_3)_2$.2.5D$_2$O \cite{xu2000,tennant2002} have been constructed which appear to conform
 to this simple model, while at the same time  more powerful neutron scattering facilities are coming
 on-line to explore
their properties. For a review and further references, see Barnes
{\it et al.} \cite{barnes1999}. On the theoretical side, Uhrig and
Schulz \cite{uhrig1996} used a field theory approach to predict
the appearance of both singlet ($S=0$) and triplet
 ($S=1$)
bound states below the two-triplet continuum. This was confirmed by later studies
\cite{fledderjohann1997,bouzerar1998,shevchenko1999a}.
Bouzerar and Sil \cite{bouzerar1998a} and Shevchenko {\it et al.} \cite{shevchenko1999a} have
treated the AHC using the Brueckner approach; while Singh and Zheng \cite{singh1999}, Trebst {\it
et al.} \cite{trebst2000} and Zheng {\it et al.} \cite{zheng2003} have carried out high-order dimer
series expansions for the model, which give an accurate numerical picture of the
dimerized phase.

In Section II, we lay out the triplet-wave expansion formalism for the case of the alternating chain. In Section III,
the expansion to leading orders on powers of the coupling $\lambda$ is discussed for the ground state energy and
energy gap. In Section IV, numerical results are presented for the ground-state energy, the one-particle spectrum, the
two-triplon bound states, and the exclusive structure factors for these states. A summary and conclusions are
presented in Section V.

\section{ Triplet-wave expansion}
\label{sec2}

\begin{figure}
  \includegraphics[width=0.9\linewidth]{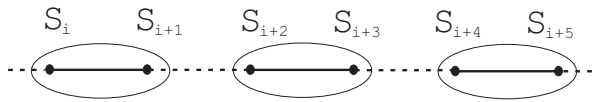}
 \caption{The alternating Heisenberg chain.}
 \label{fig1}
\end{figure}

The Hamiltonian for the alternating Heisenberg chain can be written
\begin{equation}
H = \sum_{ i \ odd} {\bf S_i \cdot S_{i+1}} + \lambda\sum_{i \ even} {\bf
S_i \cdot S_{i+1}}
\label{eq10}
\end{equation}
For $\lambda =0$, the system reduces to independent dimers as shown in
Figure \ref{fig1}. Let us consider a single dimer with two spins ${\bf S_1,S_2}$.
The four states in the Hilbert space consist of a singlet and three
triplet states with total spin $S=0,1$ respectively, and eigenvalues
\begin{eqnarray}
{\bf S_1 \cdot S_2} = \left\{ \begin{array}{cc}
-3/4 & (S=0) \\
+1/4 & (S=1)
\end{array}
\right.
\label{eq11}
\end{eqnarray}
We denote the singlet ground state as $|0>$, and introduce triplet
creation operators that create the triplet states out of the vacuum
$|0>$, as follows
\begin{eqnarray}
|0> & = & \frac{1}{\sqrt{2}}[|\uparrow \downarrow> -|\downarrow
\uparrow>] \nonumber \\
|1,x> & = & t^{\dagger}_x|0> = -\frac{1}{\sqrt{2}}[|\uparrow \uparrow> -|\downarrow
\downarrow>] \nonumber \\
|1,y> & = & t^{\dagger}_y|0> = \frac{i}{\sqrt{2}}[|\uparrow \uparrow> +|\downarrow
\downarrow>] \nonumber \\
|1,z> & = & t^{\dagger}_z|0> = \frac{1}{\sqrt{2}}[|\uparrow \downarrow> +|\downarrow
\uparrow>]
\label{eq12}
\end{eqnarray}
Then the spin operators ${\bf S_1}$ and ${\bf S_2}$ can be represented
in terms of triplet operators by
\begin{eqnarray}
S_{1\alpha} & = &
\frac{1}{2}[t^{\dagger}_{\alpha}(1-t^{\dagger}_{\gamma}t_{\gamma}) +
(1-t^{\dagger}_{\gamma}t_{\gamma}) t_{\alpha}
-i\epsilon_{\alpha\beta\gamma}t^{\dagger}_{\beta}t_{\gamma}]
 \nonumber \\
S_{2\alpha} & = &
\frac{1}{2}[-t^{\dagger}_{\alpha}(1-t^{\dagger}_{\gamma}t_{\gamma}) -
(1-t^{\dagger}_{\gamma}t_{\gamma}) t_{\alpha}
\nonumber \\
& &   -i\epsilon_{\alpha\beta\gamma}t^{\dagger}_{\beta}t_{\gamma}]
\label{eq13}
\end{eqnarray}
where $\alpha,\beta,\gamma$ take the values $x,y,z$ and repeated indices
are summed over. This is similar to the representation of Sachdev and
Bhatt \cite{sachdev1990}, except that we have omitted singlet operators
$s^{\dagger},s$, but used projection operators
$(1-t^{\dagger}_{\gamma}t_{\gamma})$ instead. Assume the triplet
operators obey bosonic commutation relations
\begin{equation}
[t_{\alpha},t^{\dagger}_{\beta}] = \delta_{\alpha\beta},
\label{eq14}
\end{equation}
then one can show that within the physical subspace (i.e. total number
of triplet states is 0 or 1), the representation (\ref{eq13}) obeys the
correct spin operator algebra
\begin{equation}
[S_{1\alpha},S_{1\beta}]  =  i\epsilon_{\alpha\beta\gamma}S_{1\gamma},
\hspace{5mm} [S_{2\alpha},S_{2\beta}] =
i\epsilon_{\alpha\beta\gamma}S_{2\gamma},
\label{eq15a}
\end{equation}
\begin{equation}
[S_{1\alpha},S_{2\beta}]  =  0
\label{eq15b}
\end{equation}
\begin{equation}
{\bf S}_1^2  =  {\bf S}_2^2 = 3/4, \hspace{5mm} {\bf S_1 \cdot S_2} =
t^{\dagger}_{\alpha}t_{\alpha} - 3/4
\label{eq15c}
\end{equation}
The projection operators ensure that we remain within the subspace.

Returning to the alternating chain, we can now define triplet operators
$t^{\dagger}_{n\alpha},t_{n\alpha}$ for each dimer $n$ along the chain.
For a chain of $N$ dimers, the Hamiltonian now can be expressed in terms
of triplet operators as
\begin{widetext}
\begin{eqnarray}
H & = &  -\frac{3N}{4} + \sum_n t^{\dagger}_{n\alpha}t_{n\alpha} -\frac{\lambda}{4}
\sum_n \{\tdna(1-\tdng \tng) \tdnoa (1- \tdnod \tnod) + (1- \tdng
\tng)\tna (1- \tdnod \tnod) \tnoa
\nonumber \\
 & & + \tdna (1- \tdng \tng) (1- \tdnod \tnod) \tnoa
+ (1- \tdng \tng) \tna \tdnoa (1- \tdnod \tnod)\}
\nonumber \\
 & & +\frac{\lambda}{4}\sum_n
\tdag_{n\be}t_{n\ca}(\tdag_{n+1,\ca}t_{n+1,\be}-\tdag_{n+1,\be}t_{n+1,\ca})
+i\frac{\lambda}{4}\epsilon_{\al\be\ca}\sum_n\{\tdag_{n\al}(
1-\tdag_{n\de}t_{n\de})\tdag_{n+1,\be}t_{n+1,\ca}
\nonumber \\
 & &
-\tdag_{n\be}t_{n\ca}
\tdag_{n+1,\al}(1-\tdag_{n+1,\de}t_{n+1.\de}) +
(1-\tdag_{n\de}t_{n\de})t_{n\al}\tdag_{n+1,\be}t_{n+1,\ca}
-\tdag_{n\be}t_{n\ca}(1-\tdag_{n+1,\de}t_{n+1,\de})t_{n+1,\al}\}
\label{eq16}
\end{eqnarray}
\end{widetext}
This expression includes terms containing up to 6 triplet operators. For
the purposes of the present calculations, we shall drop terms with more
than 4 triplet operators henceforwards.

Next, perform a Fourier transform
\begin{eqnarray}
t_{k\al} & = & (\frac{1}{N})^{1/2} \sum_n e^{ikn} t_{n\al} \nonumber \\
t^{\dagger}_{k\al} & = & (\frac{1}{N})^{1/2} \sum_n e^{-ikn} \tdag_{n\al}
\label{eq17}
\end{eqnarray}
(we set the spacing between dimers $d = 1$),
then the Hamiltonian becomes
\begin{widetext}
\begin{eqnarray}
H & = & -\frac{3N}{4} + \sum_k \tdag_{k\al} t_{k\al}  -\frac{\lambda}{4}
\sum_k\cos k [\tdag_{k\al} \tdag_{-k\al} + t_{k\al}t_{-k\al} +2\tdag_{k\al}t_{k\al}]
+ \frac{\lambda}{2\sqrt{N}}\epsilon_{\al\be\ca} \sum_{123}
\delta_{1+2-3} \sin k_1 [\tdag_{1\al} \tdag_{2\be} t_{3\ca}
\nonumber \\
 & & + \tdag_{3\ca} t_{1\al}
t_{2\be}]
 + \frac{\lambda}{2N} \sum_{1234} \{\delta_{1+2+3-4}  \tdag_{1\al}
\tdag_{2\al} \tdag_{3\ca} t_{4\ca}\cos k_1 + \delta_{1-2-3-4}
\tdag_{1\ca} t_{2\ca} t_{3\al}
t_{4\al}\cos k_4
\nonumber \\
& & +\delta_{1+2-3-4}[ \tdag_{1\al} \tdag_{2\ca} t_{3\ca}
t_{4\al}\cos k_4 + \tdag_{1\al} \tdag_{2\ca} t_{3\ca} t_{4\al}\cos k_1]\}
\nonumber \\
 & & +\frac{\lambda}{4N} \sum_{1234} \delta_{1+2-3-4}[ \tdag_{1\ca}
\tdag_{2\be} t_{3\be} t_{4\ca} - \tdag_{1\be} \tdag_{2\be} t_{3\ca}
t_{4\ca}]\cos (k_1-k_3)
\label{eq18}
\end{eqnarray}
\end{widetext}

Finally, as in a standard spin-wave analysis, we perform a Bogoliubov
transform
\begin{equation}
t_{k\al} = c_k\tau_{k\al} + s_k \tau^{\dagger}_{-k\al}
\label{eq19}
\end{equation}
where $c_k = \cosh \theta_k$, $s_k = \sinh \theta_k$, $\theta_{-k} =
\theta_k$, which preserves the boson commutation relations
\begin{equation}
[\tau_{k\al},\taud_{k'\beta}] = \delta_{kk'}\delta_{\al\be}
\label{eq20}
\end{equation}
and is intended to diagonalize the Hamiltonian up to quadratic terms.
After normal ordering, the transformed Hamiltonian up to fourth order
terms reads
\begin{equation}
H = W_0 + H_2 + H_3 + H_4.
\label{eq21}
\end{equation}
Here the constant term is
\begin{eqnarray}
W_0 & = & 3N[-\frac{1}{4}+R_2
\nonumber \\
 & & -\frac{\lambda}{2}(R_3+R_4)(1-8R_2-2R_1+R_3-R_4)]
\label{eq22}
\end{eqnarray}
expressed in terms of the momentum sums
\begin{eqnarray}
R_1 & = & \frac{1}{N} \sum_k c_k s_k \nonumber \\
R_2 & = & \frac{1}{N} \sum_k  s_k^2
\nonumber \\
R_3 & = & \frac{1}{N} \sum_k c_k s_k \cos k
\nonumber \\
R_4 & = & \frac{1}{N} \sum_k s_k^2 \cos k
\label{eq23}
\end{eqnarray}
The quadratic terms are
\begin{equation}
H_2 = \sum_{k,\al} [E_k \taud_{k\al} \tau_{k\al} +Q_k(\tau_{k\al}\tau_{-k\al} + \taud_{k\al} \taud_{-k\al})]
\label{eq24}
\end{equation}
where
\begin{widetext}
\begin{equation}
E_k  =  (c_k^2 + s_k^2)[1+4\lambda(R_3+R_4)-\frac{\lambda}{2}\cos k (1-2R_1-8R_2-2R_4)]
   - \lambda c_ks_k[\cos k (1-8R_2-2R_1+2R_3)
-2(R_3+R_4)]
\label{eq25}
\end{equation}
\begin{equation}
Q_k = c_ks_k[1-\frac{\lambda}{2}(\cos k(1-2R_1-8R_2-2R_4)-8(R_3+R_4))
 -\frac{\lambda}{4}(c_k^2+s_k^2)[\cos
k(1-8R_2-2R_1+2R_3)-2(R_3+R_4)]
\label{eq26}
\end{equation}
The third and fourth-order terms are
\begin{equation}
H_3 = \frac{\lambda}{2\sqrt{N}}\epsilon_{\al\be\ca}\sum_{123}[\delta_{1+2+3}
\Phi^{(1)}_3(\tau_{1\al}\tau_{2\be}\tau_{3\ca} +
\taud_{3\ca}\taud_{2\be}\taud_{1\al}) + \delta_{1+2-3}\Phi^{(2)}_3
(\taud_{1\al}\taud_{2\be}\tau_{3\ca} +
\taud_{3\ca}\tau_{2\be}\tau_{1\al})]
\label{eq27}
\end{equation}
and
\begin{eqnarray}
H_4 & = & \frac{\lambda}{4N}\sum_{1234}[\delta_{1+2+3+4}\Phi^{(1)}_4
(\taud_{1\al}\taud_{2\al}\taud_{3\ca}\taud_{4\ca} +
\tau_{1\al}\tau_{2\al}\tau_{3\ca}\tau_{4\ca}) +
\delta_{1+2-3-4}(\Phi^{(2)}_4
\taud_{1\al}\taud_{2\al}\tau_{3\ca}\tau_{4\ca}
+\Phi_4^{(3)}\taud_{1\al}\taud_{2\ca}\tau_{3\al}\tau_{4\ca})
\nonumber \\
& & +\delta_{1+2+3-4}\Phi^{(4)}_4
(\taud_{1\al}\taud_{2\al}\taud_{3\ca}\tau_{4\ca} +
\taud_{4\ca}\tau_{3\ca}\tau_{2\al}\tau_{1\al})]
\label{eq28}
\end{eqnarray}
\end{widetext}
where we have used the shorthand notation $1 \cdots 4$ for momenta $k_1 \cdots k_4$, and the vertex functions
$\Phi^{(i)}_3, \Phi^{(i)}_4$ are listed in Appendix A.

The condition that the off-diagonal quadratic terms vanish is
\begin{equation}
Q_k =0.
\label{eq29}
\end{equation}
In a conventional spin-wave approach, this would be implemented in leading order only, giving the condition
\begin{equation}
\tanh 2\theta_k = \frac{2c_ks_k}{c_k^2+s_k^2} = \frac{\lambda \cos k}{2[1-\lambda/2 \cos k]}
\label{eq30}
\end{equation}
This would leave some residual off-diagonal quadratic terms, arising from the normal-ordering of quartic operators. In a
`modified' approach \cite{gochev1994}, we demand that these terms vanish entirely up to the order calculated, giving the
modified condition
\begin{widetext}
\begin{equation}
\tanh 2\theta_k =  \frac{\lambda[ \cos k(1-8R_2-2R_1+2R_3)-2(R_3+R_4)]}{2[1
-\lambda \cos k(1-2R_1-8R_2-2R_4)/2+4\lambda(R_3+R_4)]}
\label{eq31}
\end{equation}
\end{widetext}
Self-consistent solutions for the N equations (\ref{eq31}), with the four parameters $R_1 \cdots R_4$ given by equation
(\ref{eq23}), can easily be found by numerical means, starting from the conventional result (\ref{eq30}).

\section{Expansion in powers of $\lambda$}
\label{sec3}

As a first check on the formalism, one may calculate the leading terms
in an expansion of the energy eigenvalues in powers of $\lambda$. From
equation (\ref{eq30}), we easily see that to order $\lambda^2$
\begin{eqnarray}
s_k & = & \frac{\lambda}{4}\cos k+\frac{\lambda^2}{8}\cos^2 k
\nonumber \\
c_k & = & 1+\frac{\lambda^2}{32}\cos^2 k
\label{eq32}
\end{eqnarray}
and hence the lattice sums (\ref{eq23}) can be evaluated
\begin{eqnarray}
R_1 & = & \frac{\lambda^2}{16}, \hspace{5mm} R_2 = \frac{\lambda^2}{32}, \nonumber \\
R_3 & = & \frac{\lambda}{8} + O(\lambda^3), \hspace{5mm} R_4 =
O(\lambda^3)
\label{eq33}
\end{eqnarray}
The leading-order behaviour of the vertex functions may easily be
deduced from Appendix A.

\begin{figure}
  \includegraphics[width=0.8\linewidth]{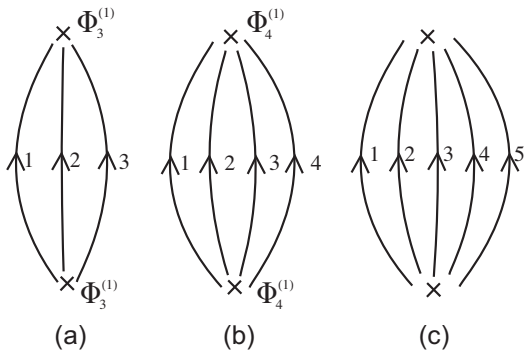}
 \caption{Perturbation diagrams contributing to the ground-state energy.}
 \label{fig2}
\end{figure}

Substituting in equation (\ref{eq22}), the ground state energy per site
is
\begin{eqnarray}
\epsilon_0 & = &  \frac{W_0}{2N} =
\frac{3}{2}\left[-\frac{1}{4}+R_2-\frac{\lambda}{2}(R_3+R_4)(1-8R_2
\right. \nonumber \\
 & & \left. -2R_1+R_3-R_4)\right]
\nonumber \\
 & \sim &
-\frac{3}{8}-\frac{3\lambda^2}{64}-\frac{3\lambda^3}{256}+O(\lambda^4),
\hspace{5mm} \lambda \rightarrow 0
\label{eq34}
\end{eqnarray}
in agreement with dimer series expansion results previously obtained for this
model \cite{singh1999}. One can easily show that perturbation diagrams
such as those in Figure \ref{fig2} do not contribute until
$O(\lambda^4)$ or higher.

The energy gap at leading order can be found from equation (\ref{eq25}):
\begin{equation}
E_k \sim 1-\frac{\lambda}{2}\cos k
+\frac{\lambda^2}{8}[4-\cos^2 k],
 \hspace{5mm} \lambda \rightarrow 0
\label{eq35}
\end{equation}

\begin{figure}
  \includegraphics[width=0.8\linewidth]{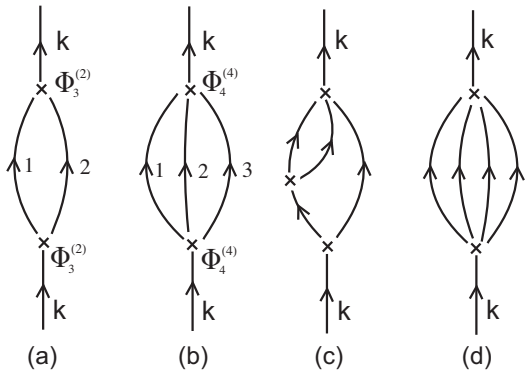}
 \caption{Perturbation diagrams contributing to the one-particle energy.}
 \label{fig3}
\end{figure}

The perturbation diagrams Figures \ref{fig3}a) and \ref{fig3}b) also
contribute at order $\lambda^2$.
Note that diagram \ref{fig3}b) does not appear in the formalism of Shevchenko
{\it et al.} \cite{shevchenko1999a}; the extra terms in our formalism are needed to implement the
hardcore constraint that two triplons cannot occupy the same site.
At leading order, the contributions of these diagrams
are
\begin{equation}
\Delta E_k^{3a)} \sim -\frac{\lambda^2}{4}(1 +
\cos k), \hspace {5mm} \lambda \rightarrow 0
\label{eq36}
\end{equation}
\begin{equation}
\Delta E_k^{3b)} \sim -\frac{\lambda^2}{4}, \hspace{5mm} \lambda \rightarrow
0
\label{eq37}
\end{equation}
(see the next section for further details). This gives a total
single-particle energy
\begin{equation}
\epsilon_k \sim 1-\frac{\lambda}{2}\cos k
-\frac{\lambda^2}{8}\cos k [2+\cos k], \hspace{5mm} \lambda \rightarrow
0
\label{eq38}
\end{equation}
which again agrees with series expansion results \cite{singh1999}.

If we compare equation (\ref{eq38}) at small momentum with the
continuum dispersion relation for a free boson,

\begin{equation}
\epsilon_k \sim \sqrt{m^2c^4 + k^2c^2}
\label{eq39}
\end{equation}
we readily discover the leading behaviour of the effective triplon
parameters, i.e. the triplon mass

\begin{equation}
m \sim \frac{2}{\lambda}[1+\lambda+O(\lambda^2)]
\label{eq40}
\end{equation}
and the `speed of light'

\begin{equation}
c \sim \frac{\lambda}{2}-\frac{3\lambda^2}{8} + O(\lambda^3)
\label{eq41}
\end{equation}
in lattice units. Note that the mass diverges and the speed of light
vanishes as $\lambda \rightarrow 0$.

\section{Numerical Results}
\label{sec4}

Writing the Hamiltonian as

\begin{equation}
H = H_0 + V
\label{eq42}
\end{equation}
where

\begin{equation}
H_0 = W_0 + H_2
\label{eq43}
\end{equation}
and
\begin{equation}
V = H_3 + H_4
\label{eq44}
\end{equation}
we can treat $H_0$ as the unperturbed Hamiltonian and $V$ as a
perturbation to obtain the leading-order corrections to the predictions
for physical quantities outlined in the previous section.
Numerical results for the model have been obtained using the
finite-lattice method. The momentum sums are carried out for a fixed
number of dimers $N$, using corresponding discrete values for the momentum
$k$, e.g.

\begin{equation}
k_n = \frac{2\pi n}{N}, \hspace{5mm}n=1. \cdots N
\label{eq45}
\end{equation}
Results were obtained for $N$ up to 40,
and a fit in powers of $1/N$ was made to extrapolate to the
bulk limit $N \rightarrow \infty$.

\subsection{Ground-state energy}

\begin{figure}
  \includegraphics[width=1.0\linewidth]{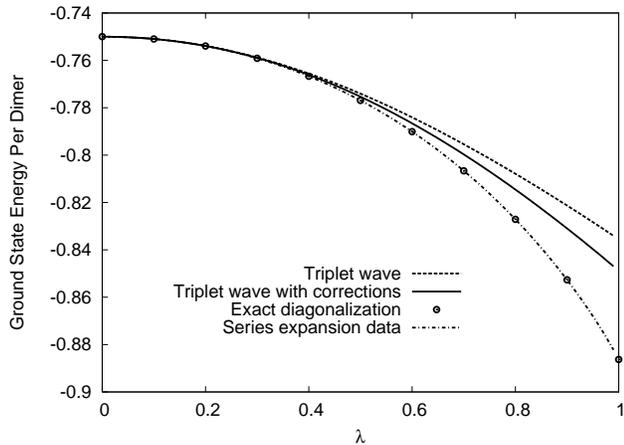}
 \caption{Ground-state energy as a function of $\lambda$.}
 \label{fig4}
\end{figure}

The leading corrections to the ground-state energy correspond to the
diagrams in Figures \ref{fig2}a) and \ref{fig2}b). Their
contributions are

\begin{widetext}
\begin{equation}
\Delta \epsilon_0^{2a)}  =  \frac{-9\lambda^2}{2N^2}\sum_{123}
\delta_{1+2+3}\frac{\Phi_3^{(1)}(123)\Phi_3^{(1)}(123)}{(E_1+E_2+E_3)}
\label{eq45a}
\end{equation}
\begin{eqnarray}
\Delta \epsilon_0^{2b)}  & = &  \frac{-3\lambda^2}{4N^3}\sum_{1234}
\delta_{1+2+3+4}\frac{\Phi_4^{(1)}(1234)}{(E_1+E_2+E_3+E_4)}
  \left[3\Phi_4^{(1)}(1234)
  +\Phi_4^{(1)}(1324)
  +\Phi_4^{(1)}(1423)\right]
\label{eq45b}
\end{eqnarray}
\end{widetext}
In leading order one can show that these terms are $O(\lambda^4)$,
whereas diagrams such as Figure \ref{fig2}c) are $O(\lambda^5)$ or
higher. The resulting bulk estimates of the ground-state energy,
including these corrections, are listed in Table \ref{tab1}. Figure
\ref{fig4} shows the behaviour of the ground-state energy as a function
of $\lambda$ resulting from this modified triplon theory, as compared
with the high-order dimer series calculations of Zheng {\it et al.}
 \cite{singh1999} and exact diagonalization data of Barnes {\it et al.}
 \cite{barnes1999}.
It can be seen that out to $\lambda \simeq 0.4$ there is
quantitative agreement between our calculation and the series
estimates, but some discrepancy emerges at larger $\lambda$.

\begin{table}
\caption{Values for the energy per dimer $\epsilon_0$ and the energy gap at $k=0$ as functions of
$\lambda$. The left-hand box giver series estimates \cite{singh1999} while the right-hand box gives our
present triplet-wave results.}
\begin{tabular}{|c|cc|cc|}
\hline
$\lambda$ &  Series & expansion & Triplet & expansion \\
 & $\epsilon_0$ & Energy gap & $\epsilon_0$ & Energy gap \\
\tableline
0.0 & -0.75000 & 1.00000 & -0.75000 & 1.00000 \\
0.1 & -0.75096 & 0.94628 & -0.75096 & 0.94647 \\
0.2 & -0.75394 & 0.88521 & -0.75392 & 0.88625 \\
0.3 & -0.75914 & 0.81684 & -0.75896 & 0.81885 \\
0.4 & -0.76672 & 0.74106 & -0.76611 & 0.74252 \\
0.5 & -0.77694 & 0.65748 & -0.77535 & 0.65483 \\
0.6 & -0.79010 & 0.56530 & -0.78659 & 0.55341 \\
0.7 & -0.80662 & 0.46300 & -0.79970 & 0.43649 \\
0.8 & -0.82712 & 0.34753 & -0.81455 & 0.30312 \\
0.9 & -0.85268 & 0.21130 & -0.83096 & 0.15309 \\
1.0 & -0.88630 & 0.00828 & -0.84878 & -0.01323 \\
\hline
\end{tabular}
\label{tab1}
\end{table}

\subsection{One-particle spectrum}

The leading corrections to the one-particle spectrum correspond to the
diagrams in Figures \ref{fig3}a) and \ref{fig3}b). Their contributions
are

\begin{widetext}
\begin{equation}
\Delta E_k^{3a)}  =  \frac{\lambda^2}{N}\sum_{{\bf 12}}\delta_{
1+2-k}\frac{\Phi_3^{(2)}(
12k)\Phi_3^{(2)}(12k)}{(E_k-E_1-E_2)}
\label{eq46}
\end{equation}
\begin{eqnarray}
\Delta E_k^{3b)} & = & \frac{\lambda^2}{8N^2}\sum_{ 123}\delta_{
1+2+3-k}\frac{\Phi_4^{(4)}(
123k)}{(E_k-E_1-E_2-E_3)}
\left[3\Phi_4^{(4)}(123k)
+\Phi_4^{(4)}(321k)
+\Phi_4^{(4)}(312k)
\right]
\label{eq47}
\end{eqnarray}
\end{widetext}
In leading order, these terms are $O(\lambda^2)$, as stated in the
previous section, while diagrams like \ref{fig3}c), d) are
$O(\lambda^3)$ or higher.

\begin{figure}
  \includegraphics[width=1.0\linewidth]{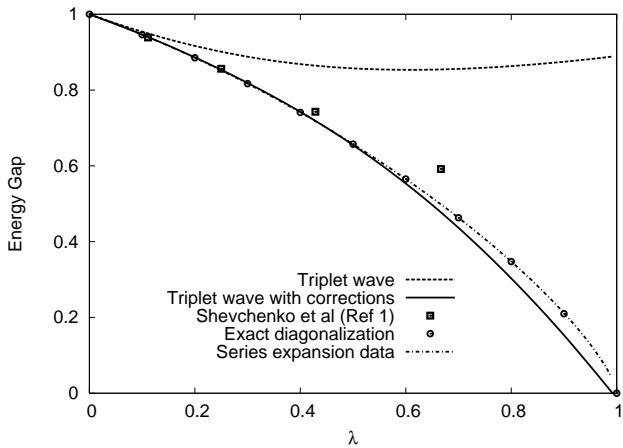}
 \caption{[Color online] Energy gap at $k=0$ as a function of $\lambda$. The dot-dashed line shows the series estimates
\cite{singh1999}, while the other lines show the leading order and
improved triplet-wave results. The filled squares are results from
Shevchenko {\it et al.} \cite{shevchenko1999a}, and circles are
results from Barnes {\it et al.} \cite{barnes1999}. }
 \label{fig5}
\end{figure}

The resulting bulk estimates of the energy gap at $k = 0$ are listed in
Table \ref{tab1}, and displayed in Figure \ref{fig5}. It can be seen
that the inclusion of the diagrams \ref{fig3}a) and \ref{fig3}b)
improves the agreement with series dramatically.
This agreement may be fortuitous, given that the
agreement for the ground-state energy is not so good, but it is
gratifying to see nevertheless. It can be seen that our present approach improves upon that of Shevchenko
{\it et al.} \cite{shevchenko1999a} at intermediate $\lambda$.

\begin{figure}
  \includegraphics[width=1.0\linewidth]{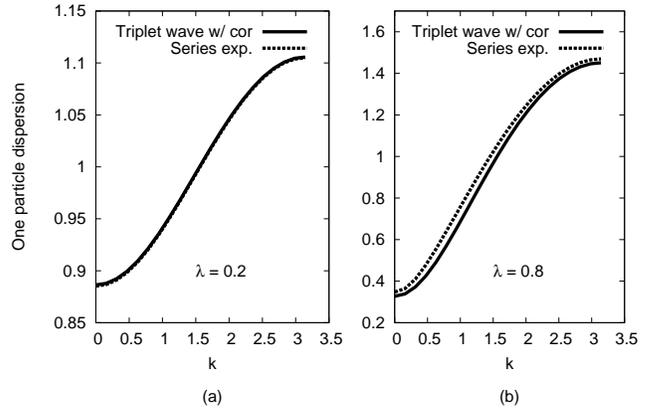}
 \caption{One-particle spectrum at selected couplings.}
 \label{fig6}
\end{figure}

\begin{figure}
 \includegraphics[width=1.1\linewidth]{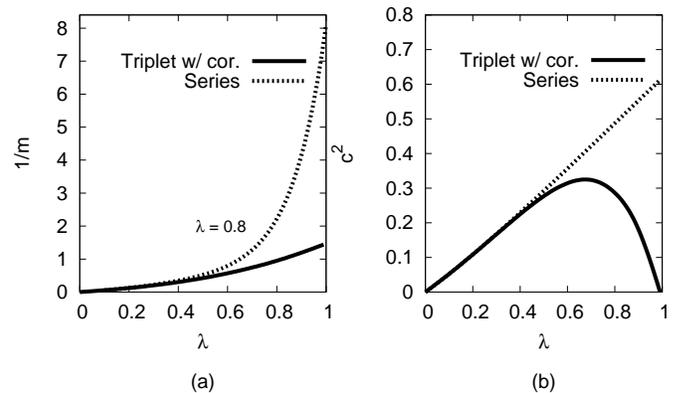}
 \caption{a) The inverse triplon mass parameter $1/m$, and b) the 'speed of light' $c^2$, as functions of $\lambda$.}
 \label{fig7}
\end{figure}

The dispersion of the one-particle energy as a function of momentum is
illustrated at selected couplings in Figure \ref{fig6}, while Figures \ref{fig7}a) and \ref{fig7}b) show the
corresponding behaviour of the inverse triplon mass parameter $1/m$ and the speed of light squared, $c^2$.
At the smaller coupling, the dispersion agrees quantitatively with series estimates, but at $\lambda =
0.8$ we can see that the minimum of the energy is too broad: the curvature at $k=0$ should diverge as
$\lambda \rightarrow 1$. This is reflected in the fact that our results for $1/m$ and $c^2$ are much too
low at large couplings. We note that the exact value of the speed of light $c$ at $\lambda=1$ is $\pi/2 =
1.57$ \cite{johnson1973}, which is about twice the value of even the series estimate ($\simeq 0.78$). This
is presumably due to the singular behaviour of the model in this limit, including logarithmic corrections,
which even high-order series expansions cannot accurately reproduce.

\subsection{Two-triplon bound states}

It has been found in previous studies \cite{uhrig1996,shevchenko1999a} that the quartic terms in
the Hamiltonian lead to attraction between two elementary triplons,
giving rise to $S=0$ and $S=1$ bound states. We look for solutions of
the two-body Schr{\" o}dinger equation

\begin{equation}
H|\psi> = E|\psi>.
\label{eq48}
\end{equation}

The two-body wave functions $|\psi(K)>$ can be written as follows:
\begin{flushleft}
{\it Singlet sector ($S=0$):}
\end{flushleft}
\begin{equation}
|\psi^S(K)> =
\frac{1}{\sqrt{6}}\sum_{q,\alpha}\psi^S(K,q)\tau^{\dagger}_{K/2+q,\alpha}
\tau^{\dagger}_{K/2-q,\alpha}|0>
\label{eq49}
\end{equation}
where $K$ is the centre-of-mass momentum and $q$ the relative momentum
of the two particles;

\begin{flushleft}
{\it Triplet sector ($S=1$):}
\end{flushleft}
\begin{equation}
|\psi^T_{\alpha}(K)> =
\frac{1}{2}\sum_{q,\beta,\gamma}\epsilon_{\alpha\beta\gamma}
\psi^T(K,q)\tau^{\dagger}_{K/2+q,\beta}\tau^{\dagger}_{K/2-q,\gamma}|0>
\label{eq50}
\end{equation}
where $K$ is the centre-of-mass momentum and $q$ the relative momentum
(we will not write out the quintuplet states explicitly).

From equation (\ref{eq48}) one can readily derive the integral
Bethe-Salpeter equation satisfied by the bound-state wave functions:

\begin{eqnarray}
[E^{S,T}(K)-E_{K/2+q}-E_{K/2-q}]\psi^{S,T}(K,q)
= \nonumber \\
\frac{1}{N}\sum_{p}M^{S,T}(K,q,p)\psi^{S,T}(K,p).
\label{eq51}
\end{eqnarray}

\begin{figure}
  \includegraphics[width=1.0\linewidth]{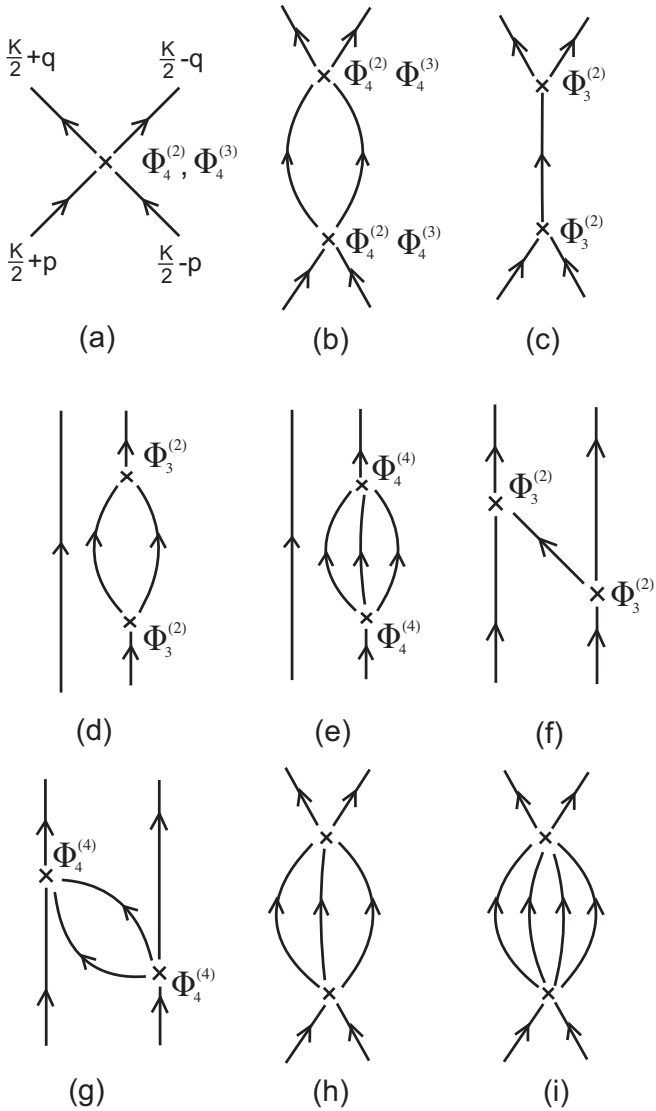}
 \caption{Perturbation diagrams contributing to the 2-particle
scattering amplitude.}
 \label{fig8}
\end{figure}

In leading order, the scattering amplitudes $M^{S,T}(K,q,p)$ are simply
given by the 4-particle vertex from the perturbation operator $V$, Figure
\ref{fig8}a). Hence we find for the different sectors:

\begin{flushleft}
{\it Singlet sector ($S=0$):}
\end{flushleft}
\begin{widetext}
%\begin{eqnarray}
\begin{equation}
M^S(K,q,p)  =
\frac{\lambda}{2}[3\Phi^{(2)}_4(K/2+p,K/2-p,K/2+q,K/2-q)
  +\Phi^{(3)+}_4(K/2+p,K/2-p,K/2+q,K/2-q)]
\label{eq52}
\end{equation}
%\end{eqnarray}
\end{widetext}
where the wave function is symmetric,
\begin{equation}
\psi^S(K,-q)=\psi^S(K,q)
\label{eq53}
\end{equation}
and the symmetric and antisymmetric pieces of the vertex function
$\Phi_4^{(3)}$ are defined:
\begin{equation}
\Phi^{(3)\pm}_4  \equiv  \frac{1}{2}[\Phi^{(3)}_4(1234)\pm
\Phi^{(3)}_4(1243)]
\label{eq54}
\end{equation}

\begin{flushleft}
{\it Triplet sector ($S=1$):}
\end{flushleft}
\begin{equation}
M^T(K,q,p) =
\frac{\lambda}{2}\Phi^{(3)-}_4(K/2+p,K/2-p,K/2+q,K/2-q)
\label{eq55}
\end{equation}
with the wave function antisymmetric
\begin{equation}
\psi^T(K,-q)=-\psi^T(K,q).
\label{eq56}
\end{equation}

\begin{flushleft}
{\it Quintuplet sector ($S=2$):}
\end{flushleft}
\begin{equation}
M^Q(K,q,p) =
\frac{\lambda}{2}\Phi^{(3)+}_4(K/2+q,K/2-q,K/2+p,K/2-p)
\label{eq57}
\end{equation}
where the wave function is once again symmetric
\begin{equation}
\psi^Q(K,-q)=\psi^Q(K,q).
\label{eq58}
\end{equation}

At leading order in $\lambda$, we find
\begin{equation}
M^S(K,q,p) \sim \lambda[\cos (K/2)(\cos p + \cos
q)-\cos p \cos q]
\label{eq56a}
\end{equation}
and
\begin{equation}
M^T(K,q,p) \sim -\frac{\lambda}{2}\sin p \sin q
\label{eq56b}
\end{equation}

Following Shevchenko {\it et al.} \cite{shevchenko1999a}, one can then find simple solutions  (unnormalized) to the
Schr{\" o}dinger equation (\ref{eq51}):
\begin{eqnarray}
\psi^S(K,q) & \sim & \frac{\cos(K/2) -\cos q}{1+\cos^2(K/2)-2\cos(K/2)\cos q},
\nonumber \\
\psi^T(K,q) & \sim & \frac{\sin q}{1+4\cos^2(K/2) -4\cos(K/2) \cos q},
\label{eq56c}
\end{eqnarray}
corresponding to bound-state energies
\begin{eqnarray}
E^S(K) & \sim & 2-\frac{\lambda}{2}(1+\cos^2(K/2))
\nonumber \\
E^T(K) & \sim & 2-\frac{\lambda}{4}(1+4\cos^2(K/2))
\label{eq56d}
\end{eqnarray}
compared to the lower edge of the 2-particle continuum
\begin{equation}
E_2(K) \sim 2 -\lambda \cos(K/2).
\label{eq56e}
\end{equation}
These agree with the dimer series expansion \cite{barnes1999,trebst2000} at leading order.

Note, however, that the singlet solution is valid for all $K$, touching the continuum at $K=0$; while the
triplet solution is only physically valid over a finite range of momenta $K=\{2\pi/3,4\pi/3\}$, and not at
$K=0$. The end-points of this range are just where the triplet bound state enters the continuum, and the
denominator of (\ref{eq56c}) for $\psi^T$ vanishes at $q=0$. Note also that both dispersion curves meet the lower edge
of the continuum at a tangent.

\begin{figure}
  \includegraphics[width=1.0\linewidth]{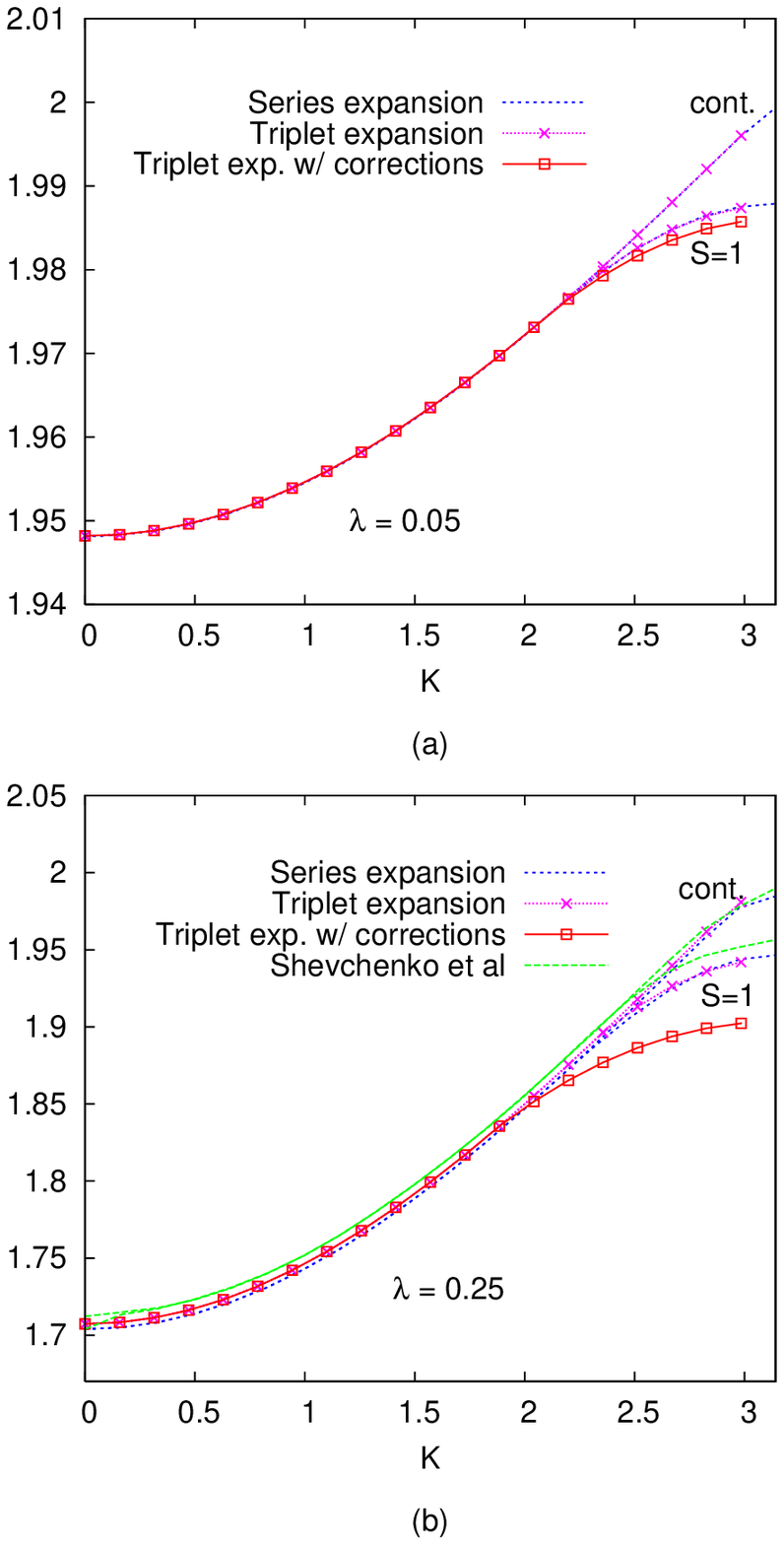}
 \caption{Dispersion relation for the triplet ($S=1$) 2-particle bound state at selected couplings a) $\lambda=0.05$ and b)
$\lambda=0.25$. The lower edge of the continuum is labelled 'cont'.}
 \label{fig9}
\end{figure}
\begin{figure}
  \includegraphics[width=1.0\linewidth]{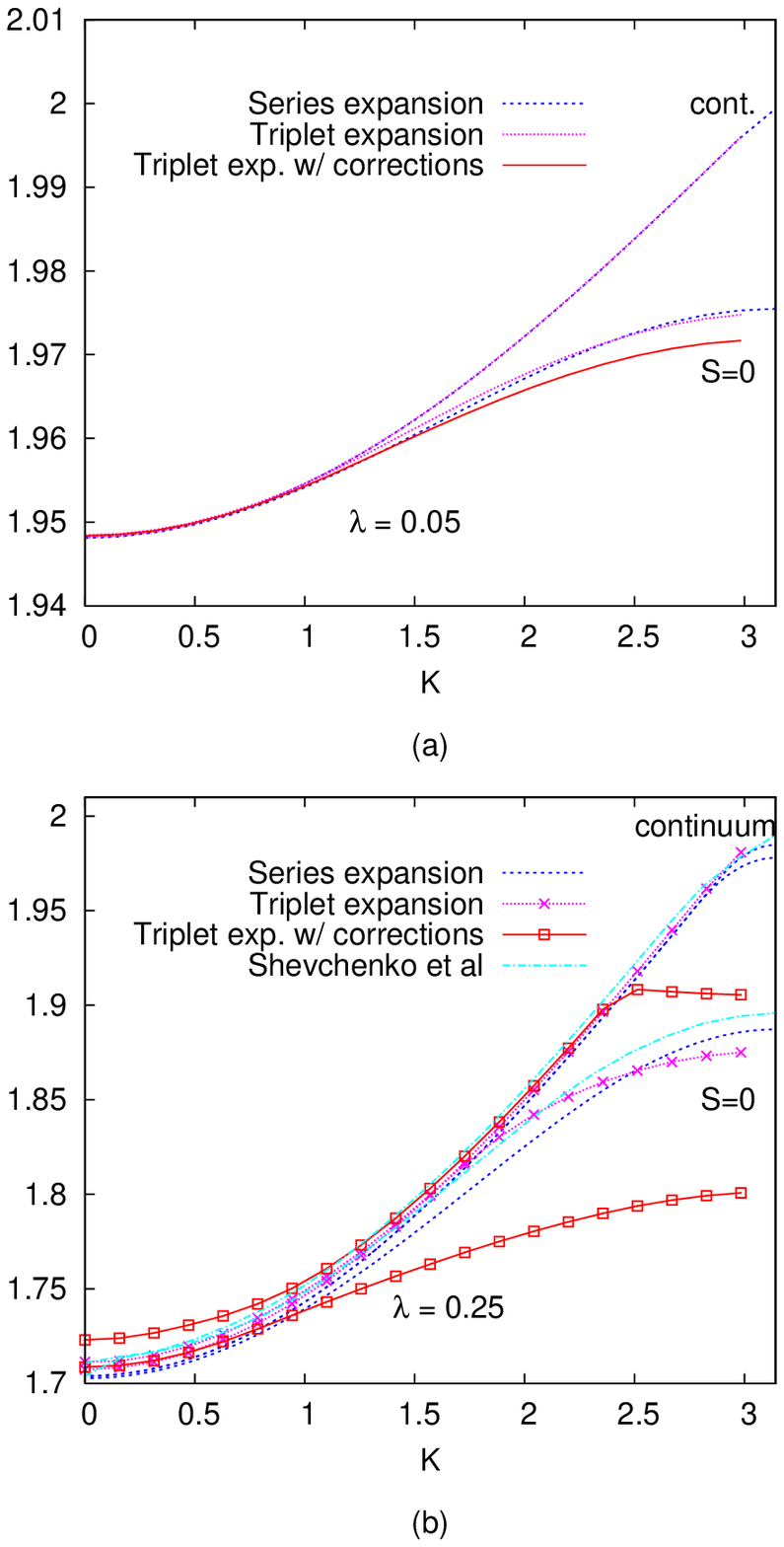}
 \caption{[Color online] Dispersion relation for singlet ($S=0$) 2-particle bound states at selected couplings a) $\lambda=0.05$ and b)
$\lambda=0.25$. The lower edge of the continuum is labelled 'cont'.}
 \label{fig10}
\end{figure}
At the next order O($\lambda^2$), further diagrams contribute, as given
in Figures \ref{fig8}b)-i). Two of these, Figures \ref{fig8}h) and
\ref{fig8}i), we are not in a position to calculate, because they
involve 5 or 6-particle vertices. Figure \ref{fig8}b) is already
accounted for by diagonalizing the effective Hamiltonian in the
2-particle subspace. Diagrams \ref{fig8}d) and \ref{fig8}e) simply
correspond to renormalizations of the single-particle energies in the
diagonal terms of the effective Hamiltonian in the 2-body sector. Finally, we
 can calculate the contribution of Figures \ref{fig8}c), \ref{fig8}f)
and \ref{fig8}g) to the effective Hamiltonian
using perturbation theory.
In general, the change in the energy eigenvalue is

\begin{equation}
\Delta E = \frac{1}{N}\sum_{p,q}\Delta M(K,q,p)\psi(K,p)\psi(K,q)
\label{eq59}
\end{equation}
where the vertex function $\Delta M(K,q,p)$ for each different diagram
and spin state is listed in Appendix B.

 The corrections due to these
diagrams can now be calculated.
On a finite lattice, equation (\ref{eq51}) becomes a matrix eigenvalue
equation, which can readily be solved numerically. We have calculated
results for lattices of up to $N=40 $.
The resulting bound-state
spectrum is displayed in Figures \ref{fig9} and \ref{fig10}.
The first thing to note is that the modified but uncorrected triplet expansion agrees with series
expansion estimates quite well, for both the lowest-lying singlet and triplet bound states. For the triplet
state, the result is
substantially better than that of
Shevchenko {\it et al.} \cite{shevchenko1999a} at $\lambda =
0.25$. Inclusion of the perturbation corrections actually makes the agreement worse, and gives much
too large binding energies, especially at $\lambda = 0.25$. This can be attributed to the neglect of
diagrams \ref{fig8}h),i), which are of the same order as the diagrams we have calculated. Unless the
extra diagrams are included, we cannot do better than the uncorrected estimates.

We have also looked for signs of
the second singlet and second triplet bound states which were found
to appear
at order $\lambda^2$
by Trebst {\it et al.} \cite{trebst2000}.
In the corrected results, a second singlet bound state does appear, in fact, but with much too large a
binding energy once again.
The
detailed dynamics of the bound states are sensitive to higher-order terms.

\subsection{Structure Factors}

The ``reduced exclusive structure factor" or spectral weight for a
specific intermediate state $\Lambda$ with momentum ${\bf K}$ can be
written

\begin{equation}
S^{\alpha\alpha}_{\Lambda}(K) = |\Omega_{\Lambda}^{\alpha}(K)|^2
\label{eq60}
\end{equation}
where
\begin{equation}
\Omega_{\Lambda}^{\alpha}(K) = \sqrt{N}\sum_{i^*}<\Psi_{\Lambda}(
K)|S^{\alpha}_{i^*}|\Psi_0>\exp (-i K \cdot r_{i^*})
\label{eq61}
\end{equation}
and the sum $i^*$ runs over sites of the unit cell on the lattice, and
N is the number of unit cells (dimers). Using equations (\ref{eq13}),
(\ref{eq17}), and (\ref{eq19}), the spin operators $S_1$ and $S_2$ on
sites 1 and 2 can be expressed in terms of triplet operators (taking $n
= 0$ in equation (\ref{eq17})):
\begin{widetext}
\begin{eqnarray}
S_{1,2}^{\alpha} & = & \pm \sum_k
T_1^{(1)}(k)(\tau_{k\alpha}+\tau^{\dagger}_{k\alpha}) -i\epsilon_{\alpha
\beta \gamma} \sum_{12} [
T_2^{(1)}(12)(\tau_{1\beta}^{\dagger}\tau_{2\gamma}^{\dagger}-\tau_{2\beta}^
{\dagger}\tau_{1\gamma}^{\dagger}+\tau_{2\beta}\tau_{1\gamma}-\tau_{1\beta}
\tau_{2\gamma})
\nonumber \\
 & & +T_2^{(2)}(12)(\tau_{1\beta}^{\dagger}\tau_{2\gamma}+
\tau_{2\beta}^{\dagger}\tau_{1\gamma}-\tau_{2\gamma}^{\dagger}\tau_{1\beta}-
\tau_{1\gamma}^{\dagger}\tau_{2\beta})]
\nonumber \\
 & & \mp \sum_{123} \left[
T_3^{(1)}(123)(\tau^{\dagger}_{1\alpha}\tau^{\dagger}_{2\gamma}\tau^{\dagger}_{3\gamma}
+\tau^{\dagger}_{2\gamma}\tau^{\dagger}_{3\gamma}\tau_{1\alpha}
+\tau^{\dagger}_{1\alpha}\tau_{2\gamma}\tau_{3\gamma}
+\tau_{2\gamma}\tau_{3\gamma}\tau_{1\alpha} )
+T_3^{(2)}(123)(\tau^{\dagger}_{1\alpha}\tau^{\dagger}_{2\gamma}\tau_{3\gamma}
 +\tau^{\dagger}_{3\gamma}\tau_{2\gamma}\tau_{1\alpha})\right]
\end{eqnarray}
\end{widetext}
where the upper and lower signs correspond to
$S_1^{\alpha},S_2^{\alpha}$ respectively, and
\begin{equation}
T_1^{(1)}(k) = \frac{1}{2\sqrt{N}}(c_k + s_k)(1-R_1-4R_2)
\end{equation}
\begin{equation}
T_2^{(1)}(12) = \frac{1}{8N}(c_1s_2-s_1c_2)
\end{equation}
\begin{equation}
T_2^{(2)}(12) = \frac{1}{8N}(c_1c_2-s_1s_2)
\end{equation}

\begin{equation}
T_3^{(1)}(123) =
\frac{(c_1 + s_1)}{4N^{3/2}}(c_2s_3+s_2c_3)
\end{equation}
and
\begin{equation}
T_3^{(2)}(123) =
\frac{(c_1 + s_1)}{2N^{3/2}}(c_2c_3+s_2s_3)
\end{equation}

\begin{figure}
  \includegraphics[width=1.0\linewidth]{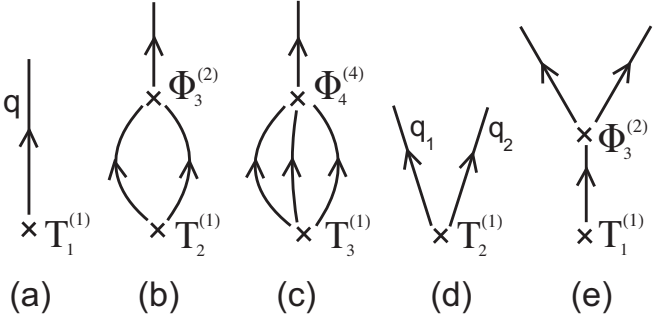}
 \caption{Perturbation diagrams contributing to exclusive structure
factors. }
 \label{fig11}
\end{figure}

In leading order
(Figure \ref{fig11}a), the one-particle matrix element is
\begin{eqnarray}
\Omega_{\Lambda}^{\alpha}(K)
 & = & i\sin(\frac{Ka}{2})[(1-R_1-4R_2)(c_K+s_K)]
\nonumber \\
 & \sim & i\sin(\frac{Ka}{2})[1+\frac{\lambda}{4}\cos K
\nonumber \\
 & & +\frac{\lambda^2}{64}(5\cos
2K -7)].
\end{eqnarray}
Here
$a$ represents the spacing between spins in the dimer, i.e. $a=1/2$ for
the uniform lattice in our present units.

Higher-order diagrams such as Figs. \ref{fig11}b), c) do not contribute
until O($\lambda^2$). Their contributions are listed in Appendix C.
Hence we find
\begin{equation}
\Omega_{1p}^{\alpha(11b)}
  \sim i\cos (\frac{Ka}{2})\frac{\lambda^2}{8}\sin K
\end{equation}
\begin{equation}
\Omega_{1p}^{\alpha(11c)}
  \sim i\sin(\frac{Ka}{2})\frac{\lambda^2}{8}.
\end{equation}
We must also account for the renormalization of the 1-particle wave
function due to Figures \ref{fig3}a) and \ref{fig3}b), giving a
multiplicative renormalization factor

\begin{widetext}
\begin{eqnarray}
 Z_K & = &
1-\frac{\lambda^2}{2N}\sum_{12}\delta_{1+2-K}\left[\frac{\Phi_3^{(2)}(12k)}{E_K-E_1-E_2}\right]^2
-\frac{\lambda^2}{16N^2}\sum_{123}\delta_{1+2+3-K}\frac{\Phi_4^{(4)}(123K)}{(E_K-E_1-E_2-E_3)^2}[3\Phi_4^{(4)}(123K)
\nonumber \\
 & &+\Phi_4^{(4)}(321K)+\Phi_4^{(4)}(312K)]
 \nonumber \\
 &  \sim & 1-\frac{\lambda^2}{8}\cos K -\frac{3\lambda^2}{16}
\end{eqnarray}
giving a total amplitude
\begin{eqnarray}
\Omega_{1p}^{\alpha}(K)
 &  \sim  & i[\sin(\frac{Ka}{2})[1+\frac{\lambda}{4}\cos K
 +\frac{\lambda^2}{64}
(-11-8\cos K +5\cos 2K)]+\cos(\frac{Ka}{2})\frac{\lambda^2}{8}\sin
K]
\end{eqnarray}
whereas Zheng et
al. \cite{zheng2003} obtain
\begin{eqnarray}
\Omega_{1p}^{\alpha}(K)
 &  \sim  & i[\sin(\frac{Ka}{2})[1+\frac{\lambda}{4}\cos K
 +\frac{\lambda^2}{6}
(-11-4\cos K +5\cos 2K)]+\cos(\frac{Ka}{2})\frac{\lambda^2}{8}\sin
K]
\end{eqnarray}

\end{widetext}
We have been unable to resolve the source of the
discrepancy at order $\lambda^2$.

The calculated results for the one-particle spectral weight are displayed in Figure 12. It can be seen
that the results match the series estimates quite well, even at larger $\lambda$. The corrected
estimates are a little low at small $k$, and a little high at large $k$: this reflects the discrepancy
at order $\lambda^2$ referred to above.

\begin{figure*}
  \includegraphics[width=0.7\linewidth]{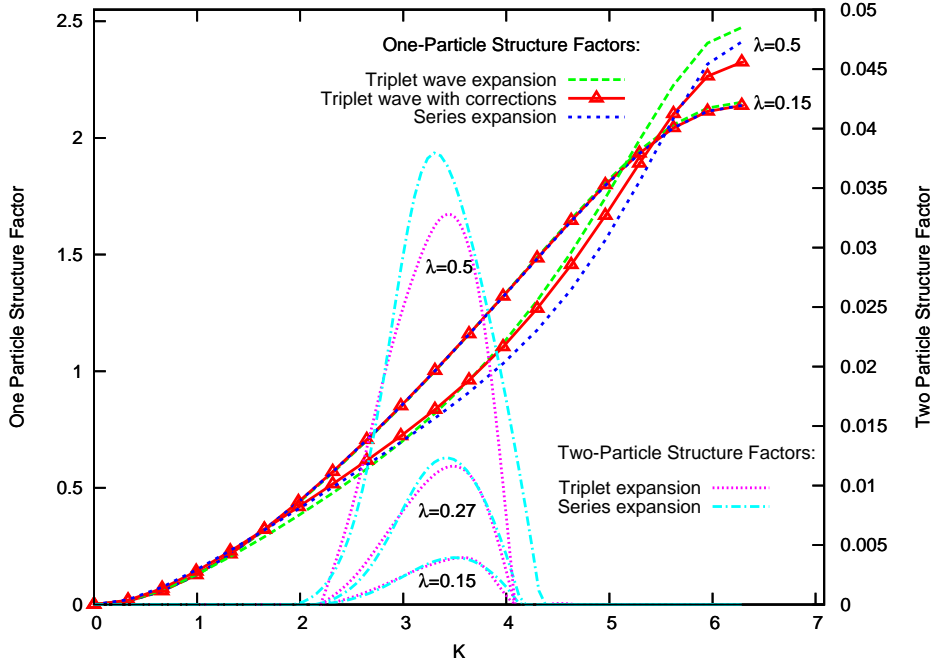}
 \caption{[Color online] The 1-particle (left axis) and 2-particle (right axis) spectral weights as functions of momentum
at selected couplings $\lambda$. }
 \label{fig12}
\end{figure*}

For the 2-particle bound states, the leading  perturbation diagrams contributing to
the exclusive structure factors are illustrated in Figs.
\ref{fig11}d),e). Their contributions for the triplet states are
\begin{eqnarray}
\Omega_T^{\alpha[11d]}(K) & = & -8i\sqrt{N}\cos (\frac{Ka}{2}) \times
\nonumber \\
 & &  \sum_{q}\psi^T(K,q)T_2^{(1)}(K/2+q,K/2-q)
\nonumber \\
 & \sim & -i\frac{\lambda}{2\sqrt{N}}\cos (\frac{Ka}{2})\sin (K/2)
\nonumber \\
 & & \times  \sum_q  \sin q \psi^T(K,q)
\end{eqnarray}
and
\begin{eqnarray}
\Omega_T^{\alpha[11e]}(K) & = & 2i\lambda \sin (\frac{Ka}{2})T_1^{(1)}(K) \times
\nonumber \\
 & & \sum_{q}\psi^T(K,q)\frac{\Phi_3^{(2)}(K/2+q,K/2-q,K)}
{E_{K/2+q}+E_{K/2-q}-E_K}
\nonumber \\
 & \sim & i\frac{\lambda}{\sqrt{N}}\sin (\frac{Ka}{2}) \cos (K/2)
\nonumber \\
 & & \times \sum_q  \sin q \psi^T(K,q)
\end{eqnarray}
Inserting the wave function \ref{eq56c}, we obtain the leading order behaviour of
the triplet bound state contribution to the structure factor (for the uniform
case $a=1/2$):
\begin{equation}
S_T^{\alpha\alpha}(K) \sim \frac{\lambda^2}{2}\sin^6(K/4)
(1-4\cos^2(K/2)),
\label{eq79}
\end{equation}
which agrees with the leading order series calculation \cite{zheng2003}. Thus
the bound-state spectral weight
vanishes at the threshold points $\{\cos(K/2) = \pm 1/2\}$ where the bound state
merges with the continuum.

The numerically calculated results for the spectral weights are displayed in Figure 12. For the 1-particle weight, it
can be seen that the corrected triplet-wave expansion matches the series estimates very well at the two
lower couplings, and only begins to deviate significantly at $\lambda=0.5$. The triplet wave expansion also
works surprisingly well for the 2-particle weight, which is only of order $1\%$ of the 1-particle weight.

\section{Summary and Conclusions}
\label{sec5}

In this paper, we have developed a modified triplet-wave expansion method for dimerized spin systems, analogous to the
modified spin-wave formalism \cite{takahashi1987,gochev1994} for magnetically ordered systems. It differs from the
earlier approaches of Sachdev and Bhatt \cite{sachdev1990} and Kotov {\it et al.} \cite{kotov1998} in that projection
operators are used to confine the system to the physical subspace in the bosonic formulation, eliminating the need for
a separate constraint. The two-body boson operators are also fully diagonalized through the highest order calculated.

The formalism has been applied to the case of the alternating Heisenberg spin chain.
Using perturbation theory to second order, we have calculated the ground-state energy per dimer, the dispersion relations
for one-particle states and two-particle bound states, and the spectral weights for these states.
 It has been shown that the
results reproduce the leading order terms in a dimer series expansion in powers of $\lambda$
\cite{singh1999,trebst2000,zheng2003}, apart from an unexplained discrepancy at order $\lambda^2$ in the 1-particle
spectral weight. The results are quantitatively accurate at small $\lambda$, but begin to show significant
discrepancies from high-order series expansions at larger $\lambda$, as one would expect.
The discrepancies become more serious for the more sensitive dynamical quantities such as two-particle binding
energies.
The inclusion of a partial set of
higher-order corrections
 for the two-particle binding energies
made things worse rather than better, as one perhaps should have expected: all terms of a similar
order in $\lambda$ must be included simultaneously if a good result is to be obtained. Nevertheless, the qualitative behaviour is correctly reproduced by the formalism. In
particular, the formation of two-triplon bound states near $K=\pi/2$ in both the singlet ($S=0$) and triplet ($S=1$)
channels, which was discovered previously \cite{uhrig1996,shevchenko1999a} is reproduced.

The behaviour of the triplet bound state near the threshold where
it merges with the continuum is interesting. We have seen that the
bound-state dispersion curve merges at a tangent to the continuum,
and that the spectral weight vanishes at the threshold. The
bound-state solution does not extend into the continuum, but
terminates at the threshold. This provides a neat example of the
phenomenon of ``quasiparticle breakdown" discussed recently in the
literature \cite{masuda2006,stone2006,zhitomirsky2006}: i.e. the
termination of a single-particle state where it enters the
continuum for one-dimensional systems.

Our results appear to be more accurate and reliable at
intermediate couplings $\lambda$ than those of Shevchenko {\it et
al.} \cite{shevchenko1999a}. However, they cannot match the
quantitative accuracy of the high-order dimer series expansions
\cite{singh1999,trebst2000,zheng2003} or exact diagonalization on
large lattices \cite{barnes1999}. The calculations could be pushed
to higher orders with the aid of a computer, but it is doubtful
whether this is worthwhile. The main value of a `lattice
bosonization' approach such as this is to provide a better
analytic understanding of the behaviour of the model, and a
half-way house towards a continuum `effective field theory' for
the model. For instance, we have shown how the triplon mass
parameter and the `speed of light' can be calculated, which would
be fundamental parameters of the effective field theory. It would
be interesting to apply the approach to dimerized models in two
dimensions.

\acknowledgments
We would like to thank Profs. J. Oitmaa and O. Sushkov
for very useful discussions and advice.
This work forms part of a research project supported by a grant
from the Australian Research Council.

\begin{widetext}
\appendix{}
\begin{center}
{\bf APPENDIX A}
\end{center}

The vertex functions $\Phi^{(i)}_3,\Phi^{(i)}_4$ are:
%Psi 1

\begin{eqnarray}
\Phi^{(1)}_3(123) & = &  \frac{1}{6}[
\sin k_1 (c_1+s_1)(c_2s_3-s_2c_3)+\sin k_2
(c_2+s_2)(c_3s_1-s_3c_1) \nonumber \\
 & & +\sin k_3 (c_3+s_3)(c_1s_2-s_1c_2)]
\label{eqA1}
\end{eqnarray}
\begin{eqnarray}
\Phi^{(2)}_3(123) & = & \frac{1}{2}[
\sin k_1(c_1+s_1)(c_2c_3-s_2s_3)+\sin k_2(c_2+s_2)(s_1s_3-c_1c_3)
 +\sin k_3(c_3+s_3)(s_1c_2-c_1s_2)]
\label{eqA2}
\end{eqnarray}
\begin{eqnarray}
\Phi^{(1)}_4(1234) & = &  \frac{1}{4}[(\cos
k_1+\cos k_2)(c_1c_2+c_1s_2+s_1c_2+s_1s_2)(c_3s_4+s_3c_4)
\nonumber \\
 & &+(\cos k_3+\cos k_4)(c_3c_4+c_3s_4+s_3c_4+s_3s_4)(c_1s_2+s_1c_2)
\nonumber \\
 & & +\cos(k_1+k_4))(c_1s_4-s_1c_4)(s_2c_3-c_2s_3)
\nonumber \\
 & & +\cos (k_1+k_3)
(c_1s_3-s_1c_3)(s_2c_4-c_2s_4)]
\label{eqA4}
\end{eqnarray}
\begin{eqnarray}
\Phi^{(2)}_4(1234) & = & \frac{1}{2}[(\cos k_1+\cos
k_2)(c_1c_2+c_1s_2+s_1c_2+s_1s_2)(s_3c_4+c_3s_4)
\nonumber \\
 & & +(\cos k_4+\cos k_3)(c_1s_2+s_1c_2)(c_3c_4+c_3s_4+s_3c_4+s_3s_4)
\nonumber \\
 & &
+\cos (k_1-k_3)(s_1c_2s_3c_4+c_1s_2c_3s_4-c_1c_2c_3c_4-s_1s_2s_3s_4)
\nonumber \\
 & &
  +\cos (k_1-k_4)(c_1s_2s_3c_4+s_1c_2c_3s_4-c_1c_2c_3c_4-s_1s_2s_3s_4)]
\label{eqA6}
\end{eqnarray}
\begin{eqnarray}
\Phi^{(3)}_4(1234) & = &
\cos k_1((c_1c_2+s_1c_2)(s_3c_4+c_3c_4)+(c_1s_2+s_1s_2)(s_3s_4+c_3s_4))
+\cos k_2((c_1c_2+c_1s_2)(c_3s_4+c_3c_4) \nonumber \\
 & & +(s_1c_2+s_1s_2)(s_3s_4+s_3c_4))
+\cos k_3((c_1s_2+s_1s_2)(c_3s_4+s_3s_4)+(s_1c_2+c_1c_2)(s_3c_4+c_3c_4))
\nonumber \\
 & & +\cos k_4((s_1c_2+s_1s_2)(s_3c_4+s_3s_4)+(c_1s_2+c_1c_2)(c_3s_4+c_3c_4))
\nonumber \\
 & & +\cos (k_1-k_4)(c_1c_4-s_1s_4)(c_2c_3-s_2s_3)
\nonumber \\
& & +\cos (k_1+k_2)(c_1s_2-s_1c_2)(c_3s_4-s_3c_4)
\label{eqA7}
\end{eqnarray}
\begin{eqnarray}
\Phi^{(4)}_4(1234) & = &
(\cos k_1 +\cos k_2)(c_1c_2+s_1s_2+c_1s_2+s_1c_2)(c_3c_4+s_3s_4)
+(\cos k_3 +\cos k_4)(c_1s_2+s_1c_2)(c_3s_4 \nonumber \\
 & & +s_3c_4+c_3c_4+s_3s_4)
+\cos (k_2+k_3)(c_1s_2c_3c_4+s_1c_2s_3s_4-c_1c_2s_3c_4-s_1s_2c_3s_4)
\nonumber \\
 & & +\cos (k_1+k_3)(s_1c_2c_3c_4+c_1s_2s_3s_4-c_1c_2s_3c_4-s_1s_2c_3s_4)
\label{eqA8}
\end{eqnarray}

We have `symmetrized' these expressions with respect to their indices,
using momentum conservation.

\begin{center}
{\bf APPENDIX B}
\end{center}

The two-body terms $\Delta M(K,p,q)$ defined in equation (\ref{eq54}) for
the diagrams Figs. \ref{fig8}c), f) and g) are as follows (the energy
denominators are `symmetrized' between initial and final states):

\begin{flushleft}
{\it Scalar state}
\end{flushleft}

\begin{equation}
\Delta M_S^{(9c)}(K,q,p) = 0
\end{equation}

\begin{eqnarray}
\Delta M_S^{(9f)}(K,q,p) & = & \lambda^2\left\{
\frac{\Phi_3^{(2)}(p-q,K/2+q,K/2+p)\Phi_3^{(2)}(K/2-p,p-q,K/2-q)}{(E_{p-q}+1/2(E_{K/2+q}+E_{K/2-p}-E_{K/2+p}-E_{K/2-q}))}
\right.
\nonumber \\
 & & \left. + (ditto, p \rightarrow -p) +(ditto, q \rightarrow -q) + (ditto, p \leftrightarrow -p, q \leftrightarrow -q)\right\}
\end{eqnarray}

\begin{eqnarray}
\Delta M_S^{(9g)}(K,q,p) & = & -\frac{\lambda^2}{16N}\sum_k\left\{
[\Phi_4^{(4)}(p-q-k,k,K/2+q,K/2+p)(3\Phi_4^{(4)}(k,p-q-k,K/2-p,K/2-q)
\right.
\nonumber \\
 & & +\Phi_4^{(4)}(K/2-p,k,p-q-k,K/2-q)+\Phi_4^{(4)}(K/2-p,p-q-k,k,K/2-q))
\nonumber \\ & &
+2\Phi_4^{(4)}(K/2+q,p-q-k,k,K/2+p)
  \times (3\Phi_4^{(4)}(K/2-p,k,p-q-k,K/2-q)
\nonumber \\
 & &
  +\Phi_4^{(4)}(k,p-q-k,K/2-p,K/2-q)
  +\Phi_4^{(4)}(K/2-p,p-q-k,k,K/2-q))]
\nonumber \\
 & &/(E_k+E_{p-q-k}+1/2(E_{K/2+q}+E_{K/2-p}-E_{K/2+p}-E_{K/2-q}))
\nonumber \\
 & &  + (ditto, p \leftrightarrow -p)
+ (ditto, q \leftrightarrow -q)
+ (ditto, p \leftrightarrow -p, q \leftrightarrow -q)\large\}
\end{eqnarray}

\begin{flushleft}
{\it Triplet state}
\end{flushleft}

\begin{equation}
\Delta M_T^{(9c)}(K,q,p) =
\lambda^2\frac{\Phi_3^{(2)}(K/2+p,K/2-p,K)
\Phi_3^{(2)}(K/2+q,K/2-q,K)}
{(E_K-1/2(E_{K/2+q}+E_{K/2-q}+E_{K/2+p}+E_{K/2-p}))}
\end{equation}

\begin{eqnarray}
\Delta M_T^{(9f)}(K,q,p) & = & \frac{\lambda^2}{2}\left\{
\frac{\Phi_3^{(2)}(p-q,K/2+q,K/2+p)\Phi_3^{(2)}(K/2-p,p-q,K/2-q)}
{(E_{p-q}+1/2(E_{K/2+q}+E_{K/2-p}-E_{K/2+p}-E_{K/2-q}))}
\right.
\nonumber \\
 & & \left. - (ditto, p \leftrightarrow -p)
- (ditto, q \leftrightarrow -q)
+ (ditto, p \leftrightarrow -p, q \leftrightarrow -q)\right\}
\end{eqnarray}

\begin{eqnarray}
\Delta M_T^{(9g)}(K,q,p) & = & -\frac{\lambda^2}{16N}\sum_k\left\{
[\Phi_4^{(4)}(p-q-k,k,K/2+q,K/2+p)(3\Phi_4^{(4)}(k,p-q-k,K/2-p,K/2-q)
\right.
\nonumber \\
 & & +
\Phi_4^{(4)}(k,K/2-p,p-q-k,K/2-q)+\Phi_4^{(4)}(p-q-k,K/2-p,k,K/2-q))
\nonumber \\
 & & +2\Phi_4^{(4)}(p-q-k,K/2+q,k,K/2+p)
 \times (\Phi_4^{(4)}(k,p-q-k,K/2-p,K/2-q)
\nonumber \\
 & & -\Phi_4^{(4)}(p-q-k,K/2-p,k,K/2-q))]
\nonumber \\
& & /(E_k+E_{p-q-k}+1/2(E_{K/2+q}+E_{K/2-p}-E_{K/2+p}-E_{K/2-q}))
\nonumber \\
 & & \left. - (ditto, p \leftrightarrow -p)
- (ditto, q \leftrightarrow -q)
+ (ditto, p \leftrightarrow -p, q \leftrightarrow -q)\right\}
\end{eqnarray}

\begin{center}
{\bf APPENDIX C}
\end{center}
Contributions to the 1-particle matrix elements $\Omega_{\Lambda}^{\alpha}(K)$ from
the diagrams shown in Figure \ref{fig11}b),c) are:

\begin{eqnarray}
\Omega_{1p}^{\alpha[11b]}(K) &= & 4i\lambda \cos
(\frac{Ka}{2})\sum_{12}\delta_{1+2-K}\frac{T_2^{(1)}(12)}
{E_K-E_1-E_2}[\Phi_3^{(2)}(21K)-\Phi_3^{(2)}(12K)]
\end{eqnarray}
\begin{eqnarray}
\Omega_{1p}^{\alpha[11c]}(K) &= & -i\frac{\lambda}{\sqrt{N}}\sin
(\frac{Ka}{2})\sum_{123}\delta_{1+2+3-K}\frac{T_3^{(1)}(123)}
{E_K-E_1-E_2-E_3}[3\Phi_4^{(4)}(321K)+\Phi_4^{(4)}(312K)+\Phi_4^{(4)}(213K)]
\end{eqnarray}

\end{widetext}


\begin{references}
\bibitem{shastry1981} B.S. Shastry and B. Sutherland, Physica {\bf
108B}, 1069 (1981).
\bibitem{lhuillier2001} For a review, see C. Lhuillier and G. Misguich,
cond-mat/0109146, Lecture Notes at the Carg{\` e}se Summer School on {\it Trends in high magnetic field science} (May
2001).
\bibitem{read1991} N. Read and S. Sachdev, Phys. Rev. Lett. {\bf 66}, 1773 (1991); {\it
ibid} {\bf 62}, 1694 (1989); G. Murthy and S. Sachdev, Nucl. Phys. B{\bf
344}, 557 (1990).
\bibitem{kotov1999} V.N. Kotov, J. Oitmaa, O.P. Sushkov and Zheng W-H., Phys. Rev. B {\bf
60}, 14613 (1999); O.P. Sushkov, J. Oitmaa and W-H. Zheng, Phys. Rev.
B{\bf 63}, 104420 (2001).
\bibitem{anderson1987} P.W. Anderson, Science {\bf 235}, 1196 (1987).
\bibitem{capriotti2003} L. Capriotti and S. Sorella, Phys. Rev. Lett.
{\bf 84}, 3173 (2000); L. Capriotti, F. Becca, A. Parola and S. Sorella,
Phys. Rev. B{\bf 67}, 212402 (2003).
\bibitem{sachdev1990} S. Sachdev and R. Bhatt, Phys. Rev. B{\bf41}, 9323
(1990).
\bibitem{chubukov1989} A.V. Chubukov, JETP Lett. {\bf 49}, 129 (1989).
\bibitem{mattis1981} D.C. Mattis, {\it The Theory of Magnetism}, Vol. II, Vol. 55
of Springer Series in Solid-State Sciences (Springer-Verlag, Berlin, 1981).
\bibitem{kotov1998} V.N. Kotov, Zheng W-H, O.P. Sushkov and J. Oitmaa, Phys. Rev.
Letts. {\bf 80}, 5790 (1998).
\bibitem{shevchenko1999} P.V. Shevchenko and O.P. Sushkov, Phys. Rev. B{\bf 59}, 8383 (1999).
\bibitem{sushkov1998} O.P. Sushkov and V.N. Kotov, Phys. Rev. Letts. {\bf 81}, 1941 (1998).
\bibitem{kotov1999a} V.N. Kotov, O.P. Sushkov and R. Eder, Phys. Rev. {\bf B59}, 6266 (1999).
\bibitem{shevchenko1999a} P.V. Shevchenko and O.P. Sushkov,
Phys. Rev. B{\bf 59}, 8383 (1999).
\bibitem{takahashi1987} M. Takahashi, Phys. Rev. Lett. {\bf 58}, 168 (1987); Phys. Rev. B{\bf 40}, 2494 (1989).
\bibitem{gochev1994} I. G. Gochev, Phys. Rev. B{\bf 49}, 9594 (1994).
\bibitem{xu2000} G. Xu, C. Broholm, D.H. Reich and M.A. Adams, Phys. Rev. Lett. {\bf 84}, 4465
(2000).
\bibitem{tennant2002} D.A. Tennant, C. Broholm, D.H. Reich, S.E. Nagler, G.E. Granroth, T. Barnes,
K. Damle, G. Xu, Y. Chen and B.C. Sales, Phys. Rev. B{\bf 67}, 054414 (2003).
\bibitem{barnes1999} T. Barnes, J. Riera and D.A. Tennant, Phys. Rev. B{\bf 59}, 11384 (1999).
\bibitem{uhrig1996} G.S. Uhrig and H.J. Schulz, Phys. Rev. B{\bf 54},
R9624 (1996).
\bibitem{fledderjohann1997} A. Fledderjohann and C. Gros, Europhys. Lett. {\bf 37}, 189 (1997).
\bibitem{bouzerar1998} G. Bouzerar, A.P. Kampf and G.I. Japaridze, Phys. Rev. B{\bf 58}, 3117
(1998)
\bibitem{bouzerar1998a} G. Bouzerar and S. Sil, Int. J. Mod. Phys. {\bf 15}, 2821 (2001).
\bibitem{singh1999} R.R.P. Singh and Zheng W-H., Phys. Rev. B{\bf 59},
9911 (1999).
\bibitem{trebst2000} S.Trebst, H. Monien, C.J. Hamer, W-H Zheng and R.R.P. Singh, Phys. Rev. Let.t
{\bf 85}, 4373 (2000); W-H Zheng, C.J. Hamer, R.R.P. Singh, S. Trebst and H. Monien, Phys. Rev.
B{\bf 63}, 144411 (2001).
\bibitem{zheng2003} W-H. Zheng, C.J. Hamer and R.R.P. Singh, Phys. Rev. Lett. {\bf 91}, 037206 (2003);
C.J. Hamer, W-H. Zheng and R.R.P. Singh, Phys. Rev.
B{\bf 68}, 214408 (2003).
\bibitem{johnson1973} J.D. Johnson, S. Krinsky and B.M. McCoy, Phys. Rev. A{\bf 8}, 2526 (1973).
\bibitem{masuda2006} T. Masuda, A. Zheludev, H. Manaka, L-P. Regnault, J-H.
Chung and Y. Qiu, Phys. Rev. Letts. {\bf 96}, 047210 (2006).
\bibitem{stone2006} M.B. Stone, I.A. Zalisnyak, T. Hong, C.L. Broholm and D.H.
Reich, Nature {\bf 440}, 187 (2006).
\bibitem{zhitomirsky2006} M.E. Zhitomirsky, Phys. Rev. B{\bf 73}, 100404R
(2006).







\end{references}
\end{document}